\documentclass[12pt]{iopart}
\usepackage{iopams}
\usepackage{graphicx}

\begin{document}

\title[Kochen-Specker Vectors]{Kochen-Specker Vectors}

\author{Mladen Pavi\v ci\'c\dag\footnote[5]{E-mail: pavicic@grad.hr},
Jean-Pierre Merlet\ddag\footnote[8]{E-mail:
Jean-Pierre.Merlet@sophia.inria.fr},
Brendan McKay\S\footnote[4]{E-mail: bdm@cs.anu.edu.au}, and
Norman D.~Megill*\footnote[6]{E-mail: nm@alum.mit.edu}
}

\address{\dag\
University of Zagreb,
Gradjevinski fakultet, POB 217, HR-10001 Zagreb, Croatia.}

\address{\ddag\ INRIA, projet COPRIN:
06902 Sophia Antipolis Cedex, France.}

\address{\S\ Department of Computer Science,
Australian National University,
Canberra, ACT, 0200, Australia.}

\address{*\ Boston Information Group, 30 Church St.,
Belmont MA 02478, U.~S.~A.}

\begin{abstract}
We give a constructive and exhaustive definition of
Kochen-Specker (KS) vectors in a Hilbert space of any
dimension as well as of all the remaining vectors of the
space. KS vectors are elements of any set of
orthonormal states, i.e., vectors in $n$-dim Hilbert
space, ${\cal H}^n$, $n\ge 3$ to which it is impossible
to assign 1s and 0s in such a way that no two mutually
orthogonal vectors from the set are both assigned 1
and that not all mutually orthogonal vectors are
assigned 0. Our constructive definition
of such KS vectors is based on algorithms that
generate MMP diagrams corresponding to blocks of
orthogonal vectors in ${\mathbb R}^n$, on algorithms that
single out those diagrams on which algebraic \hbox to 15pt{\tt 0-1}
states cannot be defined, and on algorithms that solve nonlinear
equations describing the orthogonalities of the vectors by means of
statistically polynomially complex interval analysis and
self-teaching programs. The algorithms are limited neither by the
number of dimensions nor by the number of vectors. To demonstrate
the power of the algorithms, all 4-dim KS vector systems
containing up to 24 vectors were generated and described,
all 3-dim vector systems containing up to 30 vectors were scanned,
and several general properties of KS vectors were found.
\end{abstract}


\submitto{\JPA}

\maketitle

\section{Introduction}
\label{sec:intro}

Recently proposed experimental tests of the Kochen-Specker (KS)
theorem \cite{cabell-garc98,simon-zeil00},
skepticism on the feasibility of such experiments
[3$\,$-$\,$7],
positive experiments recently carried out \cite{ks-exp},
and recent theoretical elaborations on the theorem
[9$\,$-$\,$22]
prompted a renewed interest in the KS theorem.

The KS theorem proves that there is a set of
measurements that can be carried out on a finite dimensional
quantum system in such a way that if one assumed that
the values of measured observables are completely
independent of all other observables that can be measured
on the same system, then one would run into a contradiction.
Hence, a quantum system cannot posses a definite value of a
measurable property prior to measurement, and quantum measurements
(essentially detector clicks) carried out on quantum systems
cannot be ascribed predetermined values (say 0 and 1).
To arrive at the claim, one considers an orthonormal set
of states $\{\psi_1,\dots,\psi_n\}$, i.e., vectors in $n$-dim
Hilbert space, ${\cal H}^n$, $n\ge 3$. Projectors onto these states
satisfy: $\sum_{i=1}^nP_i=I$, where $P_i=\psi_i\psi_i^\dagger$.
Now, Kochen and Specker proved \cite{koch-speck} that
there is no function $f:{\cal H}\to {\mathbb R}$ satisfying the
Sum Rule $\sum_{i=1}^nf(P_i)=f(\sum_{i=1}^nP_i)=f(I)$
for all sets of projectors $P_i$. Hence, there is at least one
set of projectors $\{P_i,P_i',\dots\}$ and the corresponding set
of vectors $\{\psi_i,\psi_i',\dots\}$ for which the Sum Rule
is not satisfied. {Choosing $f(P_i)\in \{0,1\}$
($f(I)=1$), the theorem amounts to the following claim:
In ${\cal H}^n$, $n\ge 3$, it is impossible to assign 1s and 0s
to all vectors from such a set---which we call a
{\em KS set}---in such a way that \cite{zimba-penrose}:
\begin{enumerate}
\item No two orthogonal vectors are both
assigned the value 1;
\item In any subset of $n$ mutually orthogonal vectors, not all of
the vectors are assigned the value 0.
\end{enumerate}

\smallskip
All the vectors from a KS set, as defined above, we call {\em KS
vectors}. KS vectors in each KS set form subsets of $n$ mutually
orthogonal vectors. We arrive at one subset from another
by a series of rotation in 2-dim planes around  ($n\!-\!2$)-dim
subspaces as explained in Sec.~\ref{sec:ks}. Thus, any two subsets
share at least one vector which is orthogonal to all other
vectors in both subsets and in an $n$-dim space, two subsets can
share up to $n\!-\!2$ vectors.
The KS vectors correspond to the directions of the quantisation
axes of the measured eigenstates within experiments which have
no classical counterparts, and when we speak of finding KS
vectors we mean finding these directions. We stress here that it
is not our aim to give yet another proof of the KS theorem but
to determine the class of all KS vectors from an arbitrary
${\cal H}^n$ as well as the class of all {\em non-KS vectors},
i.e., vectors from\break the remaining sets of vectors from
${\cal H}^n$. By the class of non-KS vectors we mean vectors
that allow {\tt 0-1} states and that correspond to the directions
of the quantisation axes of the measured eigenstates within experiments
which do have classical counterparts and when we speak of finding non-KS
vectors we mean finding the latter directions.

The original KS theorem \cite{koch-speck} made
use of 192 (claimed 117) 3-dim  vectors. Subsequent attempts
to reduce the number of vectors gave the following minimal
results (usually called \it records\/\rm): Bub's system
contains 49 vectors (claimed 33)~\cite{bub},
Conway-Kochen's has 51 (claimed 31)~\cite[p.~114]{peres-book},
and Peres' system has 57 (claimed 33)~\cite{peres} 3-dim
vectors \footnote[1]{The reasons why Kochen-Specker's, Bub's,
Conway-Kochen's, and Peres' systems should be considered
as 192, 49, 51, and 57 and not as 117, 33, 31, and 33 vector systems,
respectively, are given in Sec.~\ref{sec:conclusion}-(xi)
in accordance with the results independently
obtained by J.-\AA. Larsson. \cite{larsson}} ;
Kernaghan's system contains 20 4-dim vectors with the smallest
loops (see the definition below) of size two \cite{kern};
Cabello's system has 18 4-dim vectors with the smallest loops of
size three \cite{cabell-est-96a}, etc.
Reducing the number of vectors is important
for devising experimental setups \cite{cabell-99},
especially so as recently a single qubit KS scheme was
formulated \cite{cabell-03} by means of
auxiliary quantum systems (ancillas) of the measuring apparatus
and subsequently connected with the original KS formulation
\cite{aravind}. On the other hand, knowing the class of all
KS vectors is important for better theoretical insight into 
quantum theory and possibly designing quantum computers.
However, no general method for constructing
sets of KS vectors has been proposed so far and the aim of
this paper is to give one. In doing so we will follow the ideas
put forward in \cite{mporl02,mplosinj03,mpgl04}.

So far, KS vectors have been constructed either by means
of partial Boolean algebras and orthomodular lattices
\cite{koch-speck,smith04,svozil-tkadlec,tkadlec},
by direct experimental proposals
\cite{cabell-garc98,simon-zeil00,cabell-99}, or by combining
rays in ${\mathbb R}^n$ \cite{bub,peres,kern}. These approaches
have two disadvantages: first, they depend on human ingenuity
to find ever new examples and ``records,'' and second, their
complexity grows exponentially with increasing numbers of
dimensions and vectors. For example, lattices of orthogonal
$n$-tuples have $2^n$ elements (Hasse diagrams)~\cite{shimony}
and, on the other hand, the complexity of nonlinear equations
describing combinations of orthogonalities also grows
exponentially.

As opposed to this, we are able to give algorithms for
generation of all the equations that have KS vectors as their
solutions and to effectively solve them (up to a reasonably
chosen number of vectors and dimensions---limited only by the
speed of today's computers) in a way that is
essentially of a statistically polynomial complexity.
We first recognise that a description of a discrete
observable measurement (e.g., spin) in ${\cal H}^n$ can be
rendered as a {\tt 0-1} measurement of the corresponding
projector along the vector in ${\mathbb R}^n$
onto which the projector projects. Hence, we deal with
orthogonal triples in ${\mathbb R}^3$, quadruples in
${\mathbb R}^4$, etc., which correspond to possible
experimental designs, and to find KS vectors means finding
such $n$-tuples in ${\mathbb R}^n$.

The orthogonalities of vectors within these $n$-tuples
can be described by nonlinear equations of type given in
Eq.~(\ref{eq:mmp-eq}) that have solutions. There are however
billions of such nonlinear equations that have no solutions
even for the smallest KS sets. And their number grows
exponentially with the increase of both the number of KS 
vectors and the dimension of their space. So, we established
a one-to-one correspondence between nonlinear equations and
graphs (MMP diagrams). We can handle graphs exponentially 
faster than nonlinear equations but there are
nevertheless billions of them. Therefore, we designed a
self-teaching generation algorithm for MMP diagrams:
graphs containing subgraphs that correspond to equations that 
cannot have a solution are not generated. This reduces the 
generation complexity to a statistically polynomial
one and the time required for obtaining the MMP diagrams
corresponding to systems of nonlinear equations with
solutions from billions of years to hours and days.

To switch back from the MMP diagrams to nonlinear equations to
solve them at this stage  would again take quite some time.
Therefore we defined the notion of an {\it algebraic
dispersion-free state} ({\tt 0-1} state) on MMP diagrams.
It turns out that only a small percentage of the obtained MMP
diagrams cannot have {\tt 0-1} states. Their direct verification
is again of exponential complexity. So, we developed algorithms
with backtracking that discard MMP diagrams with {\tt 0-1} states
and whose complexity turns out to be statistically polynomial.

The diagrams finally obtained correspond to candidate sets of
nonlinear equations that contain KS sets provided the equations
have real solutions. Algorithms for solving nonlinear
equations, such as Gr\"obner basis and homotopy,
are also mostly of at least exponential complexity and have been
tested without success on representative systems. Therefore
we designed new ones based on interval analysis and Ritt's
characteristic set calculations and were able to reduce their
complexity to a statistically polynomial one. This rounds up the 
constructive and exhaustive definition of KS sets and vectors and 
makes their generation feasible for reasonably chosen numbers of 
vectors and dimensions.

The paper is organised as follows. In Sec.~\ref{sec:diagrams},
MMP diagrams are defined and the algorithms as well as the
programs for their generation are presented. In 
Sec.~\ref{sec:states}, we give the algorithm and program for
finding whether MMP diagrams can be assigned a set of
dispersion-free {\tt 0-1} states and determining the latter 
sets when there is at least one set of {\tt 0-1} states 
together with the smallest MMP diagrams that do not allow 
{\tt 0-1} states. In Sec.~\ref{sec:ks}, we establish a link 
between MMP diagrams that do not allow {\tt 0-1} states and 
systems of nonlinear equations whose solutions are the KS 
vectors. We then give the algorithms and methods for solving 
the equations in a statistically polynomial time. In 
Sec.~\ref{sec:conclusion}, we present the new results we 
obtained.

\section{MMP diagrams}
\label{sec:diagrams}

We start by describing vectors as vertices (points)
and orthogonalities between them as edges (lines connecting
vertices), thus obtaining MMP diagrams 
\cite{mporl02,mpgl04,bdm-ndm-mp-1} which are defined as follows:
\begin{itemize}
\item[1.]Every vertex belongs to at least one edge;
\item[2.]Every edge contains at least 3 vertices;
\item[3.]Edges that intersect each other in $n-2$
         vertices contain at least $n$ vertices;
\end{itemize}

\smallskip
Isomorphism-free generation of MMP diagrams follows the general
principles established by~\cite{mckay98}, which we now recount briefly.

Deleting an edge from an MMP diagram, together with any vertices that
lie only on that edge, yields another MPP diagram (perhaps the vacuous
one with no vertices).  Consequently, every MMP diagram can
be constructed by starting with the vacuous diagram and adding one edge
at a time, at each stage having an MMP diagram.

We can represent this process as a rooted tree whose vertices correspond
to MMP diagrams whose vertices and edges have unique labels.  The vacuous
diagram is at the root of the tree, and for any other diagram its parent
node is the diagram formed by deleting the edge with the highest label.
The isomorph rejection problem is to prune this tree until it contains
just one representative of each isomorphism class of diagram.  This
can be achieved by the application of two {\em rules}.

Given a diagram $D$, we can identify the valid positions to add a new edge
such that Conditions~3--4 are enforced.  According to the symmetries
of $D$, some of these positions are equivalent.  The first rule is that
exactly one position in each equivalence class of positions is used;
a node in the tree formed by adding an edge in any other position is
deleted together with all its descendants.

To understand the second rule, consider a diagram $D'$ with at least
one edge.  We label the edges of $D'$ in a canonical order, which is an
order independent of any previous labelling.  Then we define the {\it
major class} of edges as those that are equivalent under the symmetries
of $D'$ to the edge that is last in canonical order.  The second rule
is: when $D'$ is constructed by adding an edge $e$ to a smaller diagram,
delete $D'$ (and all its descendants) unless $e$ is in the major class
of edges of~$D'$.

According to the theory in \cite{mckay98}, application of both rules
together is sufficient: exactly one diagram from each isomorphism class
remains in the tree.  Our im\-ple\-menta\-tion used {\tt nauty} 
\cite{mckay90} for computing symmetries and canonical orderings.  
The method allows for very efficient parallelisation of the 
computation. A generation tree for MMP diagrams with 9 vertices and 
the smallest loop of size 5 is shown in the Fig.~\ref{fig:mmp}.

\begin{figure*}[hbt]
\begin{center}
\includegraphics[width=\textwidth]{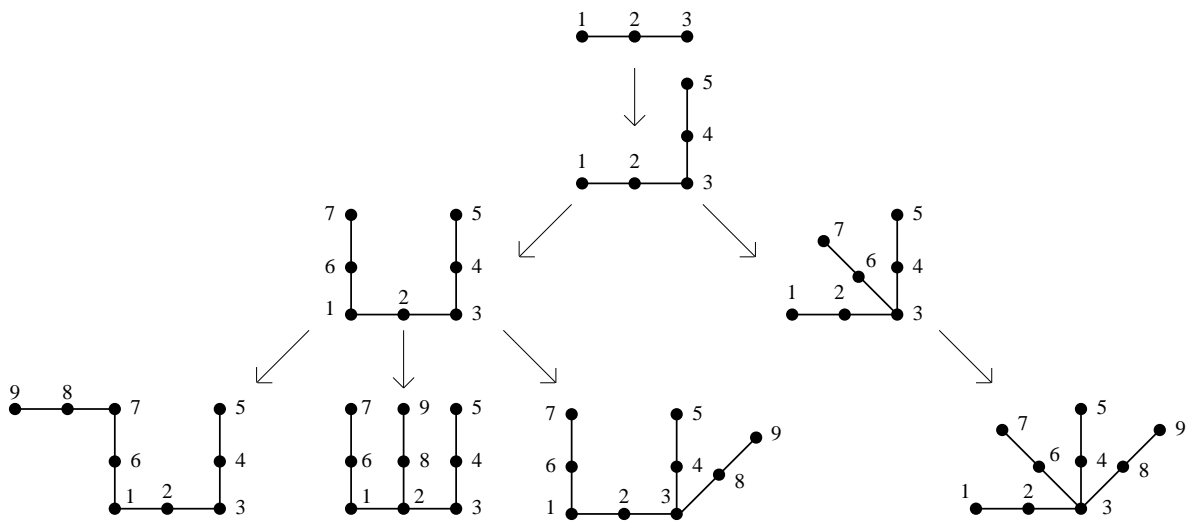}
\end{center}
\caption{An example of a generation tree for connected MMP diagrams:
9 vertices and the smallest loop of size 5 (for 9 vertices a loop cannot
be formed; the first loop appears with 10 vertices:
{\tt 123,345,567,789,9A1}). Cf.~\cite{mpgl04,bdm-ndm-mp-1}}
\label{fig:mmp}
\end{figure*}

MMP diagrams with three vertices per edge and with smallest
loops (edge polygons) of size five graphically resemble Greechie
diagrams~\cite{svozil-tkadlec}. Greechie diagrams are a
handy way to draw Hasse diagrams that represent
orthomodular lattices. The complexity of Hasse diagrams
grows exponentially with increasing dimensions and the
smallest loops of the corresponding are of size 5, while MMP
diagrams allow loops of size 4, 3, and 2 (2 edges share at
least 2 vertices). Besides, it would be quite a challenge to
find a direct lattice representation of KS vectors.

We denote vertices of MMP diagrams by {\tt 1,2,..,A,B,..a,b,..}
By the above algorithm we generate MMP diagrams with
chosen numbers of vertices and edges and a chosen minimal loop size.
E.g., in the examples given in Fig.~\ref{fig:diagrams} we
generate diagrams with 4 vertices within an edge and minimal
loops of size 2 and 3. Our programs handle diagrams with up to 90
vertices, but this limit could easily be extended.

\section{Algebraic states on MMP diagrams}
\label{sec:states}

To find diagrams that cannot be ascribed {\tt 0-1} values
we apply an algorithm which we call {\tt states01}.
The algorithm is an exhaustive search of MMP diagrams with backtracking.
The criterion for assigning {\tt 0-1} (dispersion-free) states
is that each edge must contain exactly one vertex assigned to 1, with
the others assigned to 0. As soon as a vertex on an edge is assigned a
1, all other vertices on that edge become constrained to 0, and so on.
The algorithm scans the vertices in some order, trying 0 then 1,
skipping vertices constrained by an earlier assignment.  When no
assignment becomes possible, the algorithm backtracks until all possible
assignments are exhausted (no solution) or a valid assignment is found.
In principle the algorithm is exponential, but because the diagrams of
interest are tightly coupled, constraints build up quickly.
For the range of diagram size in our study, we found that the average
time per diagram appeared to grow polynomially with the diagram size.

To implement the algorithm we wrote a program that selects
MMP diagrams with 3 and 4 vertices per edge on which {\tt 0-1} states
cannot be defined. The smallest such diagrams are given in
Fig.~\ref{fig:diagrams}.
\begin{figure*}[hbt]
\begin{center}
\includegraphics[width=\textwidth]{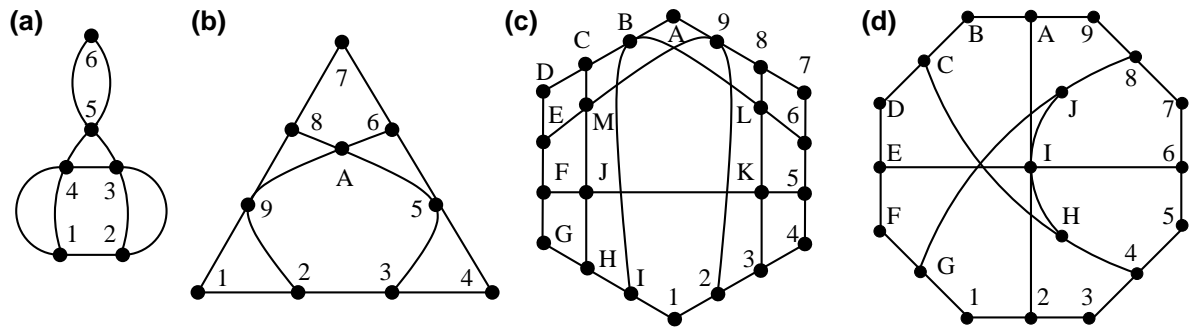}
\end{center}
\caption{Smallest MMP diagrams without {\tt 0-1} states:
(1) 4 vertices per edge: (a) loops of size 2: 6 vertices---3 edges;
(b) loops of size 3: 10-5;
(c) loops of size 4: 22-11;
(2) 3 vertices per edge: (d) loops of size 5: one of two 19-13; the other
is shown in Fig.~2$\,$(b) of \cite{mpgl04}.}
\label{fig:diagrams}
\end{figure*}

\begin{itemize}
\item {\it 3 vertices per edge}
\item[] 7 vertices---5 edges (smallest loops of size 3):
{\tt 123,345,561,275,476} (triangle);
\item[] 15-11 (4):
{\tt 123,345,567,789,9AB,BC1,CD6,2DA,2E8,4FA,CEF} (hexagon),
\item[] \qquad\qquad\ {\tt 123,345,567,789,9AB,BCD,DE1,4AE,28C,2FA,6FD}
(heptagon);
\item[] 19-13 (5):
{\tt 123,345,567,789,9AB,BCD,DEF,FG1,2IA,6IE,4HC,8JG,HIJ}~Fig.~\ref{fig:diagrams}$\,$(d),
\item[]\qquad\ \ \ \
{\tt 123,345,567,789,9AB,BCD,DE1,EI7,2F9,4GB,IJG,FJH,CH6} (heptagon);
\end{itemize}

\begin{itemize}
\item {\it 4 vertices per edge}
\item[] 6-3 (smallest loops of size 2): {\tt 1234,2356,1456} \
Fig.~\ref{fig:diagrams}$\,$(a);
\item[] 10-5 (smallest loops of size 3):  {\tt 1234,4567,7891,35A8,29A6}
Fig.~\ref{fig:diagrams}$\,$(b);
\item[] 22-11 (smallest loops of size 4):
\item[] \quad
{\tt 1234,4567,789A,ABCD,DEFG,GHI1,FJK5,HJMC,3KL8,IBL6,29ME.}
Fig.~\ref{fig:diagrams}$\,$(c),
\item[] \quad
{\tt 1234,4567,789A,ABCD,DEF1,FGH5,EMJ6,2GLC,3IJ8,HIKB,MLK9} (pentagon);
\item[] 38-19 (5):
{\tt 1234,1567,289A,5BCD,8BEF,3GHI,6JKL,GJMN,CHOP,EMQR,OQST,RUVW,
\qquad\qquad 4UXY,9SZa,FIbc,KTXb,7VZc,ALPW,DNYa} (dodecagon).
\end{itemize}

\noindent
Further details of the algorithm and the program {\tt states01}
will be given elsewhere.~\cite{pmmm04b}

\section{Kochen-Specker vectors}
\label{sec:ks}

To find KS vectors we follow the idea put forward in
\cite{mporl02,mpgl04} and proceed so as to require that
their number, i.e.~the number of vertices within edges,
corresponds to the dimension of ${\mathbb R}^n$ and that edges
correspond to $n(n-1)/2$ equations resulting from inner products
of vectors being equal to zero which means
orthogonality. So, e.g., an edge of length 4, {\tt BCDE},
represents the following 6 equations:
\begin{eqnarray}
&&{\mathbf a}_B\cdot{\mathbf a}_C=
a_{B1}a_{C1}+a_{B2}a_{C2}+a_{B3}a_{C3}+a_{B4}a_{C4}=0,\nonumber\\
&&{\mathbf a}_B\cdot{\mathbf a}_D=
a_{B1}a_{D1}+a_{B2}a_{D2}+a_{B3}a_{D3}+a_{B4}a_{D4}=0,\nonumber\\
&&{\mathbf a}_B\cdot{\mathbf a}_E=
a_{B1}a_{E1}+a_{B2}a_{E2}+a_{B3}a_{E3}+a_{B4}a_{E4}=0,\nonumber\\
&&{\mathbf a}_C\cdot{\mathbf a}_D=
a_{C1}a_{D1}+a_{C2}a_{D2}+a_{C3}a_{D3}+a_{C4}a_{D4}=0,\nonumber\\
&&{\mathbf a}_C\cdot{\mathbf a}_E=
a_{C1}a_{E1}+a_{C2}a_{E2}+a_{C3}a_{E3}+a_{C4}a_{E4}=0,\nonumber\\
&&{\mathbf a}_D\cdot{\mathbf a}_E=
a_{D1}a_{E1}+a_{D2}a_{E2}+a_{D3}a_{E3}+a_{D4}a_{E4}=0.
\label{eq:mmp-eq}
\end{eqnarray}
Each possible combination of edges for a chosen number of vertices
corresponds to a system of such nonlinear equations. A solution
to systems which correspond to MMP diagrams without {\tt 0-1}
states is a set of components of KS vectors we want to
find. Thus the main clue to finding {\em all} KS vectors is the
exhaustive generation of all MMP diagrams as given in
Sec.~\ref{sec:diagrams}, then picking out all those diagrams
that cannot have {\tt 0-1} states as presented in
Sec.~\ref{sec:states}, establishing the correspondence between
the latter diagrams and the equations for the vectors as shown
in Eq.~(\ref{eq:mmp-eq}), and finally solving the systems of
the so obtained equations.

In practice, we actually merge these four stages so as to avoid
generating those diagrams that cannot have a solution.
\footnote[8]{This merging is crucial. Without it we would not
be able to reduce the exponential complexity of the problem
to the statistically polynomial one.}
For systems of equations of type given by Eq.~(\ref{eq:mmp-eq})
that  do have solutions that do not allow {\tt 0-1} states,
such solutions are KS vectors that correspond to vertices of MMP
diagrams. Mutually orthogonal vectors correspond to edges, and
connected edges, i.e., MMP diagrams themselves correspond to
the systems of equations. For instance, in the connected
edges {\tt 1234,4567} vectors {\tt 1,2,3,4} and {\tt 4,5,6,7},
are mutually orthogonal and {\tt 4567} is obtained from
{\tt 1234} by 4-dim rotations ({\tt 1234} and {\tt 4567}
are connected by {\tt 4}). A general 2-dim rotation is
a rotation by an angle around a fixed point in the same plane.
A general 3-dim rotation is a rotation in a 2-dim plane by an
angle around a fixed axis perpendicular to this plane. So, we
define a general 4-dim rotation as a rotation in a 2-dim plane
by an angle around a fixed 2-dim plane. What is common to all
these rotations is that they always take place in a 2-dim plane.
Hence, we define an $n$-dim rotation as a rotation in a 2-dim
plane by an angle around a fixed ($n\!-\!2$)-dim subspace~\cite{duffin94}.
This also explains the case of a smallest loop of size 2 in the
4-dim case. E.g., we arrive at {\tt 4561} from  {\tt 1234}
by a rotation in the 2-dim plane determined by the vectors
{\tt 2,3} (and also by the vectors {\tt 5,6}) around the plane
determined by the vectors {\tt 1,4}.

Finding KS vectors  is not a well-posed problem in terms of
solving, though. Indeed if ${\cal V}$ is a
KS vector then $\lambda{\cal V}$ is also a KS vector for any
non-zero scalar  $\lambda$. Furthermore, if ${\cal S}$ is a
set of KS vectors, then ${\cal R}{\cal S}$ is also such a set
for any arbitrary rotation matrix ${\cal R}$. We may simplify the
problem by considering only unit vectors (i.e. vectors whose Euclidean
norm is 1 and hence vectors whose components have a value
in the range [-1,1]). To avoid the rotation
problem, we may assume that one $n$-tuple is the orthonormal basis
of ${\mathbb R}^n$. Under these assumptions, some of the orthogonality
equations simplify. E.g., if {\tt 1234} is the basis of
${\mathbb R}^4$ with {\tt 1}=$[0,0,0,1]$ then {\tt 1567} indicates
that the fourth components of {\tt 5,6,7} are 0. The
non-collinearity constraints also plays an important role. E.g.,
{\tt 1235} is not a possible $n$-tuple as three components of
{\tt 5} would be 0 and hence {\tt 5} would be collinear with {\tt 4}.

This has prompted us to develop a {\em preliminary pass}, which allows
elimination of $n$-tuples that cannot lead to a solution. Consider a
system of $m$ 4-dim vectors. The preliminary pass makes
use of an $m \times 4$ table ${\cal T}$, called the {\em 0-table},
with an entry set to 1 when a  vector component cannot be 0.
For example, if vector {\tt j} has components $[a_{j1},0,0,a_{j4}]$,
then neither $a_{j1}$ nor $a_{j4}$ can be 0 (otherwise the vector will
be collinear with one of the vectors of the basis)
and ${\cal T}[j,1]={\cal T}[j,4]=1$.
The preliminary pass selects a set of four 4-dim vectors as the basis of
${\mathbb R}^4$. It then applies a set of simplification rules on the
the orthogonality equations. For example, if the equation is
$a_{jk}~a_{ik}=0$ and ${\cal T}[j,k]=1$ then $a_{ik}$ is set to 0.
Each time a vector component value is determined the 0-table is
updated and the preliminary pass is restarted. The process will stop
when no further simplification may be performed or when a
constraint violation occurs (e.g., one equation
implies that $a_{jk}$ should be 0 while the 0-table indicates that this
component cannot be 0) in which case the system cannot have a
solution. The simplification rules used in the preliminary pass depend
on the space dimension.

The preliminary pass has been implemented as a C program that has been
added as a filter in the generation program. For avoiding
the exponential growth of the number of
generated MMP diagrams it is essential that the candidate KS-sets should be
generated incrementally i.e. that the program generates sequentially
all systems starting with a given $m$ $n$-tuples before modifying the
$m$th $n$-tuple. By using this incremental generation
during the preliminary pass determines that
an initial set of $m$ $n$-tuples has no solution and that
no further systems starting with this set will be
generated. E.g., for 18 vectors and 12 quadruples, without
such a filter we would generate $> 2.9\cdot 10^{16}$
systems---what would require more than 30 million years on a 2 GHz
CPU---while the filter reduces the generation to 100220 systems
(obtainable within $<30$ mins on a 2 GHz CPU). Thereafter
{\tt states01} gives us 26800 systems without {\tt 0-1} states
in $<5$ secs.

For the remaining systems, two solvers have been developed. One is
based on a specific implementation of Ritt characteristic set
calculation~\cite{ritt50}. Assume
that a vector {\tt V}=$[a_{V1},0,0,a_{V4}]$  (this implies
that $a_{V1}, a_{V4}$ cannot both be 0) is orthogonal to
{\tt W}=$[a_{W1},0,a_{W2},a_{W4}]$. From the
orthogonality condition $a_{W1}~a_{V1}+a_{W4}~a_{V4}=0$, we
deduce that $a_{W1}=-a_{W4}~a_{V4}/a_{V1}$ (as $a_{V1}$ cannot
be 0) and that $a_{W4}$ cannot be 0. This information
is propagated to the other equations and, as for the
preliminary pass, simplification rules are applied to the equations,
allowing us to determine further unknowns and to update the 0-table.
The process is repeated until no further unknowns can be determined
or until a constraint violation occurs. If no violation occurs, we
will usually get a set of remaining equations that is quite simple
and that allows us to determine all solutions. This solver has been
implemented using the symbolic computation software Maple. E.g.,
for system (a) in
Fig.~\ref{fig:solutions1}:
{\tt 1234,4567,789A,ABCD,DEFG,GHI1,35CE,29BI,68FH}, we get (in $<$ 10
secs on a 2 GHz CPU) the remaining set of 10 equations:
$2a_{62}^2=2a_{C1}^2=2a_{G3}^2=2a_{54}^2=4a_{E4}^2=1$, \
$2a_{53}^2=2a_{I2}^2=2a_{94}^2=4a_{A2}^2=2a_{I1}^2=1$ \
from the roots of which we may deduce the other vector components
(e.g., we get 6 as $a_{62}[0,1,-2a_{54}a_{94}a_{53},a_{94}]$).
The drawback of this approach is that it does not always allow us to
completely solve the equational systems: we may end up with a system
with fewer equations for which no further constraints can be propagated.

\begin{figure*}
\begin{center}
\includegraphics[width=\textwidth]{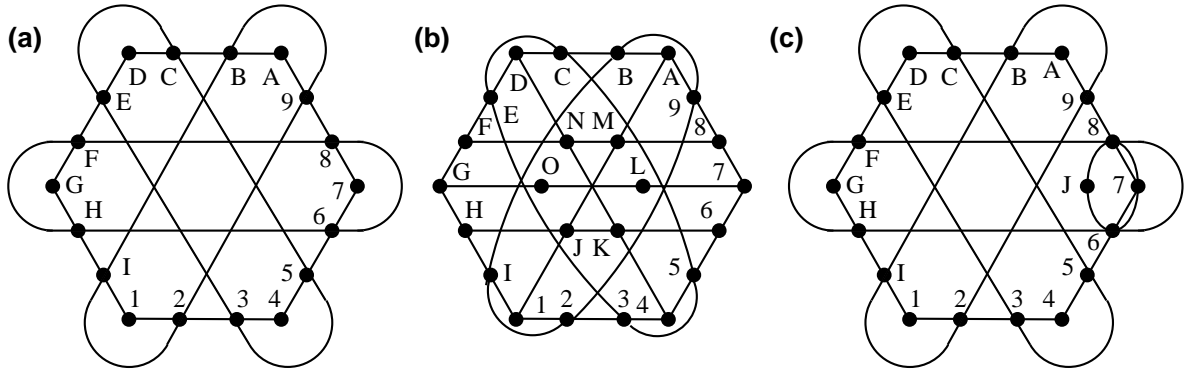}
\end{center}
\caption{Smallest 4-dim KS systems with: (1) loops of size 3: (a) 18-9
(isomorphic to Cabello et al.~\cite{cabell-est-96a}); (b) 24(22)-13
not containing system (a), with values $\not\in\{-1,0,1\}$; (2)
loops of size 2: (c)~19(18)-10.}
\label{fig:solutions1}
\end{figure*}

Our second solver is based on interval analysis. An {interval
evaluation} of an equation $f(x_1,\ldots,x_m)=0$ is a range $F=[a,b]$
such that if all the unknowns $x_1,\ldots,x_m$
are restricted to lie within given ranges, then whatever is the
values of the unknowns in their range we have $a\le f(x_1,\ldots,x_n)\le
b$. A simple way to calculate an interval evaluation is to use
interval arithmetic that simply replaces all mathematical operators
by an interval equivalent. E.g., the interval evaluation of the
orthogonality condition
$x_1y_1+x_2y_2$ with $x_1, x_2 \in [0.5,1]$ and $y_1 \in [0.1,0.2]$,
$y_2 \in
[0.2,1]$ is calculated as
$[0.5,1][0.1,0.2]+[0.5,1][0.2,1]=[0.05,0.2]+[0.01,1]=[0.06,1.2]$. Note
that if the interval evaluation of an equation does not include 0 then
there is no value of the unknowns in their range that can cancel the
equation. A {\em box} will be a set of ranges, one for each unknowns.

Solving a KS-system is an appropriate problem for interval analysis,
since all the unknowns are in the range [-1,1]. The set of these
unknowns is the box ${\cal B}_0$.
The system of equations to be solved
consists of the equations derived from the orthogonality
conditions between the vectors and of the {\em unitary equations} that
describe that each vector is a unit vector.

 A basic solver uses a list of boxes
${\cal L}$ that initially has element ${\cal B}_0$. At step $i$,
the algorithm processes box ${\cal B}_i$ of ${\cal L}$ and
calculates the interval evaluation of the orthogonality
and unitary equations: if the interval evaluation of one of these
equations does not include 0, then the algorithm will process the
next box in the list. Otherwise two new boxes will be generated
from ${\cal B}_i$ by bisecting the range of the box.
These boxes will be added to the list, and the next box in the list
will be processed. The algorithm will stop either when all the boxes
have been processed (meaning the system has no solution) or when the
width of all the ranges in a box is less than a small value while
the interval evaluations of all the equations still include 0
(meaning a solution is obtained).
Note that the method is mostly sensitive to the number of unknowns
(which explains why the vectors that appear only once in
the KS-system should be eliminated) and not so much on the number
of equations. On the contrary, additional equations may even reduce
the computation time. For example, consider a triplet
${\bf X_iX_jX_k}$ in 3D: using the orthogonality
condition, ${\bf X_k}$ is obtained as ${\bf X_k}=\pm {\bf
X_i}\times{\bf X_j}$. We get therefore two possible solutions for
${\bf X_k}$, and additional equations will be obtained by writing
that the square of each component of ${\bf X_k}$ should be equal
to the square of the same component of ${\bf X_i}\times{\bf X_j}$,

Numerous methods may be used to improve the efficiency of the basic
solver (especially to prove that indeed a system has a solution).
We use the interval analysis library {\tt
ALIAS} \footnote[2]{www.inria.fr/coprin/logiciels/ALIAS/ALIAS.html}
to deal efficiently with the KS systems.

Interval analysis has in principle an exponential complexity, due to
the bisection process. But it has been experimentally shown that in
some cases, the practical complexity is only polynomial. According
to our tests (over 400 billion systems have been checked) it
appears that the solving of the KS systems has indeed only a
statistically polynomial complexity. It must also be noted that the
solver may be used during the generation of the MMP diagrams as a
complement to the preliminary pass to avoid the exponential growth
of the number of generated MMP diagrams. Indeed, for a diagram that
has not been rejected by the preliminary pass, we may run  the
solver to further check if the diagram has a solution. But since the
solver may be relatively computer intensive, we have to use an
adaptive version in which the number of allowed bisections is
limited. For example this number may be large for relatively small
sub-graphs because determining that they don't have a solution
allows us to avoid the generation of a large number of diagrams.
On the other hand, the number of allowed bisections will be small
for sub-graphs whose size is close to the maximum (and consequently
from which few diagrams will be deduced), thus avoiding increased
generation time.

We also developed a checking program that finds solutions from
assumed sets, say $\{-1,0,1\}$, even faster ($<$ 1 sec on a
2 GHz CPU) by precomputing all possible scalar products.
The main algorithm scans the vertices and tries to assign unique
vectors to them so that all vectors assigned to a given edge are
orthogonal.  In case of a conflict the algorithm backtracks,
until either all possible assignments have been exhausted or a
solution is found. We match its exponential behaviour by scanning
next those vertices most tightly coupled to those already scanned,
helping to force conflicts to show up early on so that backtracking
can take care of them more quickly.

Further details of the
algorithms and programs presented in this section will be given
elsewhere. \cite{pmmm04b}

\section{New results and conclusions}
\label{sec:conclusion}

In this paper we presented algorithms that generate and
those that solve sets of arbitrary many Kochen-Specker
(KS) vectors that are of polynomial complexity or at least
of statistically polynomial complexity. The algorithms merge 
generation of MMP diagrams corresponding to blocks of 
orthogonal vectors in ${\mathbb R}^n$ (Sec.~\ref{sec:diagrams}), 
singling out MMP diagrams on which {\tt 0-1} states cannot be 
defined (Sec.~\ref{sec:states}), and solving nonlinear equations 
describing the orthogonalities of the vectors by means of 
interval analysis (Sec.~\ref{sec:ks}), so as to eventually 
generate KS vectors in a statistically polynomially complex way. 
Using the algorithms we obtained the following results:

(i) A general feature we found to hold for all MMP diagrams
without {\tt 0-1} states we tested is that the number of edges,
$b$ and the number of vertices that share more than one edge,
$a^*$ satisfy the following inequality: $nb\ge 2a^*$, where $n$
is the number of vertices per edge. Hence, there are no KS vectors
that share at least 2 of $\>b\>$ $n$-tuples in their KS set whose
number $a^*\> >\>\frac{n b}{2}$. In ${\mathbb R}^n$ this means that
we cannot arrive at systems with more unknowns than equations
when we disregard the unknowns that appear in only two equations.
To prove the feature for an arbitrary $n$ remains an open problem.

(ii) For MMP diagrams without {\tt 0-1} states with 3 vertices 
per edge and $a<30$ as well as with 4 vertices per edge and 
$a<23$ the stronger inequality holds: $nb\ge 2a$. The only 
exception to this rule we have found is the original 
Kochen-Specker system with 192 vertices [see (xi) and 
Fig.~\ref{fig:ks117}]. At what $a$ for a chosen $n$ this
inequality ceases to hold is an open problem. 

(iii) None of the systems corresponding to the smallest
diagrams without {\tt 0-1} states given in Sec.~\ref{sec:states}
and Fig.~\ref{fig:diagrams} has a solution. \footnote[4]{Still,
they might be significant for other fields. E.g., the two
diagrams 19-13(5) given in Sec.~\ref{sec:states} [one of them
is shown in Fig.~2$\,$(b) of \cite{mpgl04} and the other in
Fig.~\ref{fig:diagrams}$\,$(d)] are equivalent to the Greechie
diagrams with 19 {\em atoms} and  13 {\em blocks} and to our
knowledge, the smallest Greechie diagram with 3 atoms per edge
without {\tt 0-1} states known so far was the one given by
Greechie \cite{greechie71,ptak-pulm}, with 27 atoms and 18
blocks. The system 38-19(5) from Sec.~\ref{sec:states} 
yields the smallest Greechie diagram with 4 atoms per block.} 
The smallest KS vectors that we found to have real solutions 
are presented in Figs.~\ref{fig:solutions1} and 
\ref{fig:solutions2}.

(iv) Between the 4-dim system shown in 
Fig.~\ref{fig:solutions1}$\,$(a) and the one shown in 
Fig.~\ref{fig:solutions1}$\,$(a) (both with
smallest loops of size 3) there are 62 systems with loops of
size 3, all containing the system (a), 37 of which do not
have solutions from $\{-1,0,1\}$.
System (b) is the first system with loops of
size 3 not containing (a):
{\tt 1234,4567$\!$,789A,ABCD,DEFG,GHI1,FNM8,GOL7,HJK6,DNK4,AMJ1,35CE,B29J}.~One~of~its solutions is:
{\tt 12\ldots NO} =
\{1,0,1,1\}\{1,0,-2,1\}\{1,0,0,-1\}\{0,1,0,0\}\{0,0,1,0\}\{0,0,0,1\}\{1,0,0,0\}\break\{0,2,2,1\}\hfil\{0,2,-1,-2\}\hfil\{0,1,-2,2\}\hfil\{3,2,2,1\}\hfil\{1,-2,0,1\}\hfil\{-1,0,1,1\}\hfil\{1,1,0,1\}\hfil\{1,-1,1,0\}\hfil\{0,1,1,-1\}\break
\{1,1,-1,0\}\{1,-1,0,-1\}\{1,-2,-1,0\}\{1,0,1,0\}\{0,0,1,1\}\{3,2,-1,-2\}\{1,0,-1,2\}\{0,2,-1,1\} (which can, of course, easily be normalised.
The sys\-tem does not have a solution from  $\{-1,0,1\}$.

(v) The smallest 4-dim system with the smallest loop
of size 2 is the following 19-10 one:
{\tt 1234,4567,789A,ABCD,DEFG,GHI1,35CE,29BI,68FH,678I}
shown in Fig.~\ref{fig:solutions1}$\,$(c). It contains system
(a) of Fig.~\ref{fig:solutions1} and it is the only MMP
system with 19 vertices which has a solution from  $\{-1,0,1\}$
for the corresponding vectors.

(vi) The two smallest 4-dim systems with the smallest loops
of size 2 that do not contain system
(a) of Fig.~\ref{fig:solutions1} are the following 20-10
ones: {\tt 1234,4567,789A,ABCD,DEFG,GHI1,68FH,12JI,1J9B,345K,4KEC}
and {\tt 1234,4567,789A,\\ABCD,DEFG,GHI1,68FH,2IAK,345J,4JEC,9ABK}
shown in  Figs.~\ref{fig:solutions2}$\,$(a) and (b). The latter
system is isomorphic to Kernaghan's system \cite{kern}. A solution
to the former one is: {\tt 12\ldots JK} =
\{0,0,0,1\}\{1,0,0,0\}\{0,1,1,0\}\{0,1,-1,0\}\{1,0,0,-1\}\{1,1,1,1\}\{1,-1,-1,1\}\{1,1,-1,-1\}\break
\{1,0,1,0\}\hfil\{0,1,0,1\}\hfil\{1,0,-1,0\}\hfil\{1,1,1,-1\}\hfil\{1,-1,1,1\}\hfil\{1,-1,-1,-1\}\hfil\{0,0,1,-1\}\hfil\{1,1,0,0\}\hfil\{1,-1,0,0\}\break
\{0,0,1,0\}\{0,1,0,0\}\{1,0,0,1\}.

\begin{figure*}[hbt]
\begin{center}
\includegraphics[width=\textwidth]{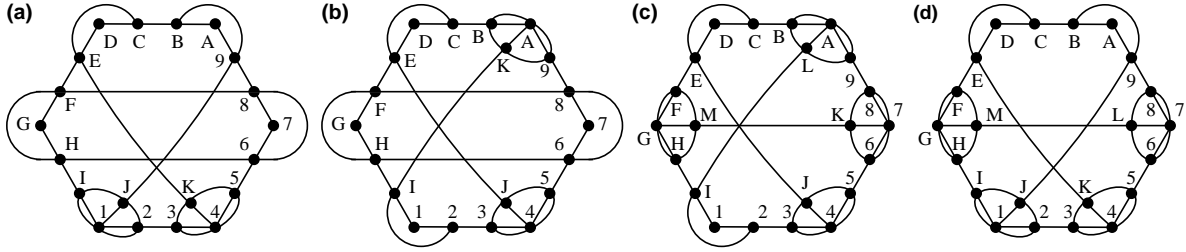}
\end{center}
\caption{Smallest 4-dim KS systems with loops of size 2:
(1) --- not containing system (a) of Fig.~\ref{fig:solutions1}: (a)
20-11; (b) 20-11 isomorphic to Kernaghan \cite{kern};
(2) --- containing neither system (a) of Fig.~\ref{fig:solutions1}
nor systems (a) and (b) off this figure: (c) 22-13; (d) 22-13.}
\label{fig:solutions2}
\end{figure*}

(vii) All 4-dim systems with up to 22 vectors and 12 edges with
the smallest loops of size 2 which do have solutions from
$\{-1,0,1\}$ contain at least one of the systems (a) and (b) of
Fig.~\ref{fig:solutions2} and in many cases also (a)
of Fig.~\ref{fig:solutions1}. The two smallest 4-dim systems
with the smallest loops of size 2 that contain neither of
the latter three systems are 22-13 systems (c) and (d) of
Fig.~\ref{fig:solutions2}:\break
{\tt 1234,\hfil 4567,\hfil 789A,\hfil ABCD,\hfil DEFG,\hfil GHI1,\hfil 2ILA,\hfil 345J,\hfil 4JEC,\hfil 678K,\hfil 7KMG,\hfil 9ABL,\hfil FGHM}
\hfil\ and\break
{\tt 1234,\hfil 4567,\hfil 789A,\hfil ABCD,\hfil DEFG,\hfil GHI1,\hfil 12IJ,\hfil 345K,\hfil 678L,\hfil GML7,\hfil 1J9B,\hfil 4KEC,\hfil FGHM}. \hfil
Their\break  solutions are: {\tt 12\ldots M}=\{1,1,0,0\}\{1,-1,0,0\}\{0,0,1,0\}\{0,0,0,1\}\{1,0,0,0\}\{0,1,1,0\}\{0,1,-1,0\}\break
\{1,0,0,1\}\{1,-1,-1,-1\}\{1,1,1,-1\}\{1,-1,1,1\}\{1,0,-1,0\}\{0,1,0,1\}\{1,0,1,0\}\{1,1,-1,-1\}\{1,-1,-1,1\}\break
\{1,-1,1,-1\}\{0,0,1,1\}\{0,1,0,0\}\{1,0,0,-1\}\{1,1,-1,1\}\{1,1,1,1\} and {\tt 12\ldots M}=\{0,0,0,1\}\{1,0,0,0\}\break
\{0,1,1,0\}\hfil\{0,1,-1,0\}\hfil\{1,0,0,-1\}\hfil\{1,1,1,1\}\hfil\{1,-1,-1,1\}\hfil\{1,-1,1,-1\}\hfil\{1,1,0,0\}\hfil\{0,0,1,1\}\hfil\{1,-1,0,0\}\break
\{1,1,1,-1\}\hfil\{1,1,-1,1\}\hfil\{1,-1,-1,-1\}\hfil\{0,1,0,-1\}\hfil\{1,0,1,0\}\hfil\{1,0,-1,0\}\hfil\{0,1,0,0\}\hfil\{0,0,1,0\}\hfil\{1,0,0,1\}\break
\{1,1,-1,-1\}\{0,1,0,1\}.

(viii)  As shown in \cite{mpgl04}, Peres' 4-dim vectors~\cite{peres}
build a hexagon with 24 vertices and 24 edges and not with 22 edges as
presented by Tkadlec in \cite{tkadlec}. (One can easily verify that
the edges \{1,-1,1,-1\}\{1,1,-1,-1\}\{1,-1,-1,1\}\{1,1,1,1\} and
\{1,-1,1,1\}\{1,-1,-1,-1\}\break \{1,1,1,-1\}\{1,1,-1,1\}
are missing in the middle of Fig.~1 in \cite{tkadlec}.) This Peres'
4-dim 24-24 KS system contains systems (a) and (c) from
Fig.~\ref{fig:solutions1} and all the systems from Fig.~\ref{fig:solutions2}.

(ix) 4-dim sys\-tems with more than 41 vectors cannot have solutions
from $\{-1,0,1\}$, and there are no such solutions to systems without
{\tt 0-1} states with minimal loops of size 5 up to 41 vectors
[there are altogether two such systems: 38-19(5) given in 
Sec.~\ref{sec:states} and a 40-20(5) system], what 
brings the Hasse (Greechie) diagram approach to the KS problem
\cite{svozil-tkadlec,tkadlec} into question.

(x) It can easily be shown that a 3-dim system of equations
representing diagrams containing loops of size 3 and 4
cannot have a real solution.
For loops of size 3, e.g.\ {\tt 123,345,561} the proof runs as
follows. Let us choose {\tt 1}=\{1,0,0\}, {\tt 2}=\{0,1,0\},
{\tt 3}=\{0,0,1\}, and {\tt i}=\{$a_{i1},a_{i2},a_{i3}$\},
$i=4,5,6$ and consider block {\tt 345}. Using
{\tt 3}$\cdot${\tt 5}=0 we get $a_{53}=0$.
Let us next consider group {\tt 561}. Using
{\tt 5}$\cdot${\tt 1}=0 we get $a_{51}=0$
Hence, {\tt 5}=\{0,$a_{52}$,0\} and is therefore collinear with
{\tt 2}. Thus, the system cannot have a solution.
The proof for loops of size 4 is similar, only a little longer.

(xi) The smallest 3-dim systems without a {\tt 0-1} valuation
have a minimal loop of size 5, 19 vertices and 13 edges
[Sec.~\ref{sec:states}, Fig.~\ref{fig:diagrams}$\,$(d)], but they
do not have real solutions. We scanned all systems with up to
30 vectors and 20 orthogonal triads and there are no KS
vectors among them. This does not mean that Conway-Kochen's
system (CK)~\cite[p.~114]{peres-book} is the smallest
KS system, though. It turns out that we cannot drop vectors
that belong to only one edge from orthogonal triads because
(a) there are cases where a solution to a full system allows
{\tt 0-1} valuation while one to a system with dropped vectors
does not and (b) there are cases where the full system does not
allow {\tt 0-1} valuation but has no solution.
So, CK is actually not a 31 but a 51 vector system:
{\tt 123,\break 145,\hfil 267,\hfil 2AB,\hfil 3CD,\hfil CEF,\hfil CGm,\hfil DIn,\hfil DKL,\hfil 6EM,\hfil 6KN,\hfil 7IO,\hfil 7GP,\hfil 4GQ,\hfil 4Ko,\hfil 5Ep,\hfil 5IS,\hfil ALW,\hfil AFX,\break
BSY,\hfil BQZ,\hfil 3cf,\hfil 3de,\hfil cOh,\hfil dMT,\hfil cN9,\hfil dP8,\hfil eSl,\hfil fQg,\hfil iR1,\hfil jk1,\hfil iFa,\hfil jLb,\hfil kOU,\hfil kMV,\hfil RPH,\hfil RNJ}\break
 with 37 edges. (Tkadlec's claims \cite{tkadlec} that CK can have
55 and 56 vectors and 54 edges are wrong.)
A solution to CK is
{\tt 12\dots op}=\{0,0,1\}\{1,0,0\}\{0,1,0\}\{1,-1,0\}\{1,1,0\}\break
\{0,1,-1\}\hfil\{0,1,1\}\hfil\{2,5,1\}\hfil\{2,5,-1\}\hfil\{0,1,2\}\hfil\{0,2,-1\}\hfil\{1,0,1\}\hfil\{1,0,-1\}\hfil\{1,-1,-1\}\hfil\{1,2,-1\}\hfil\{1,1,-1\}\break
\{2,-1,-5\}\hfil\{1,-1,1\}\hfil\{2,-1,5\}\hfil\{1,1,1\}\hfil\{1,-2,1\}\hfil\{2,1,1\}\hfil\{2,-1,-1\}\hfil\{2,1,-1\}\hfil\{2,-1,1\}\hfil\{1,1,2\}\hfil\{1,2,0\}\break
\{1,-1,-2\}\hfil\{2,-5,1\}\hfil\{2,1,5\}\hfil\{2,1,-5\}\hfil\{5,2,-1\}\hfil\{5,-2,1\}\hfil\{5,1,2\}\hfil\{5,-1,-2\}\hfil\{1,2,5\}\hfil\{1,-2,-5\}\hfil\{1,0,2\}\break
\{1,0,-2\}\hfil\{2,0,1\}\hfil
\{2,0,-1\}\hfil\{1,-5,2\}\hfil\{2,-5,-1\}\hfil\{2,-1,0\}\hfil\{2,1,0\}\hfil\{1,-2,0\}\hfil\{1,5,-2\}\hfil\{1,-2,-1\}\hfil\{1,2,1\}\break
\{1,1,-2\}\{1,-1,2\}. Thus, when all the vectors are taken into account,
Bub's system \cite{bub} with 49 vectors and 36 edges:
{\tt 123,345,167,AB6,AC4,DEG,DFH,F9O,E8V,5JI,7MN,GIa,\break
HNh,7LT,5KR,DAe,UTS,PRS,1GP,3HU,3Vj,Pgh,Uba,1Oi,VZg,OYb,6Xk,4Wn,Sde,dci,\break
dfj,imn,jlk,akQ,hnQ,eQ2} \ is so far the smallest.

\begin{figure*}[hbt]
\begin{center}
\includegraphics[width=0.8\textwidth]{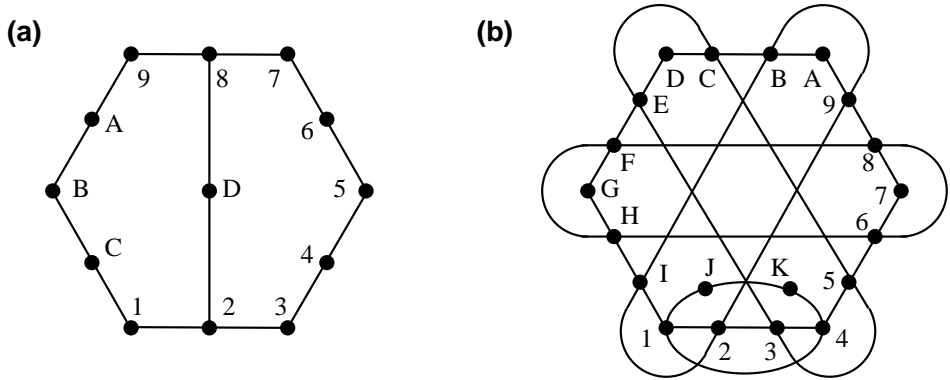}
\end{center}
\caption{(a) System with dropped vectors that belong to only one
edge ({\tt 4,6,A,C,D}) does not have any set of {\tt 0-1} states.
Taking any of these vectors into account results with systems
with at least one set of {\tt 0-1} states; (b) Neither the system
itself nor the systems obtained by dropping vectors ({\tt J} or
{\tt K} or both) allow {\tt 0-1} states. The latter systems have
solutions while the original system does not have any.}
\label{fig:drop}
\end{figure*}

Let us see why we cannot drop vectors that belong to only one edge
in detail. First, as mentioned above, if we drop all vectors that
belong  to only one edge Fig.~\ref{fig:drop}$\,$(a) we get:
{\tt 123,35,567,789,9B,B1,28}. This system has no
{\tt 0-1} states. But if we add back a single such vector,
say {\tt 123,345,567,789,9B,B1,28} or {\tt 123,35,567,789,9B,B1,2D8},
the system has at least one set of {\tt 0-1} states. All these
systems do have solutions. The opposite situation is given by
Fig.~\ref{fig:drop}$\,$(b). The system does not admit {\tt 0-1}
states but has no solution.
If we dropped {\tt J} or {\tt K} or both we would  have a
system with no {\tt 0-1} states and the systems would have solutions.
E.g., the system with dropped {\tt K} has the following solution:
{\tt 12\ldots IJ} = \{0,0,0,1\}\{1,0,0,0\}\{0,1,1,0\}\{0,1,-1,0\}\{1,0,0,-1\}\{1,-1,-1,1\}\{1,1,1,1\}\break
\{1,-1,1,-1\}\hfil\{0,1,0,-1\}\hfil\{1,0,-1,0\}\hfil\{0,1,0,1\}\hfil\{1,-1,1,1\}\hfil\{1,1,1,-1\}\hfil\{1,1,-1,1\}\hfil\{0,0,1,1\}\hfil\{1,-1,0,0\}\break\{1,1,0,0\}\{0,0,1,0\}\{1,-1,-1,0\}.
Second, in any KS diagram and therefore in Kochen-Specker
(see Fig.~\ref{fig:ks117}), Bub, Peres, and Conway-Kochen's ones
in particular, only all vectors together make a complete description
of their KS sets. 
Recall that one arrives from an
$n$-tuple to an adjoining one by rotation around an ($n\!-\!2$)-dim
subspace. E.g., in Fig.~\ref{fig:ks117}~($\!${\em ii}) one starts
with {\tt 123} and by rotations around {\tt 3}, {\tt 5}, and
{\tt 7} one arrives at {\tt 789}. So, {\tt 4} and {\tt 6} are
indispensable for the construction and cannot be dropped as
also shown by Larsson ~\cite{larsson}.

\begin{figure*}[hbt]
\begin{center}
\includegraphics[width=0.98\textwidth]{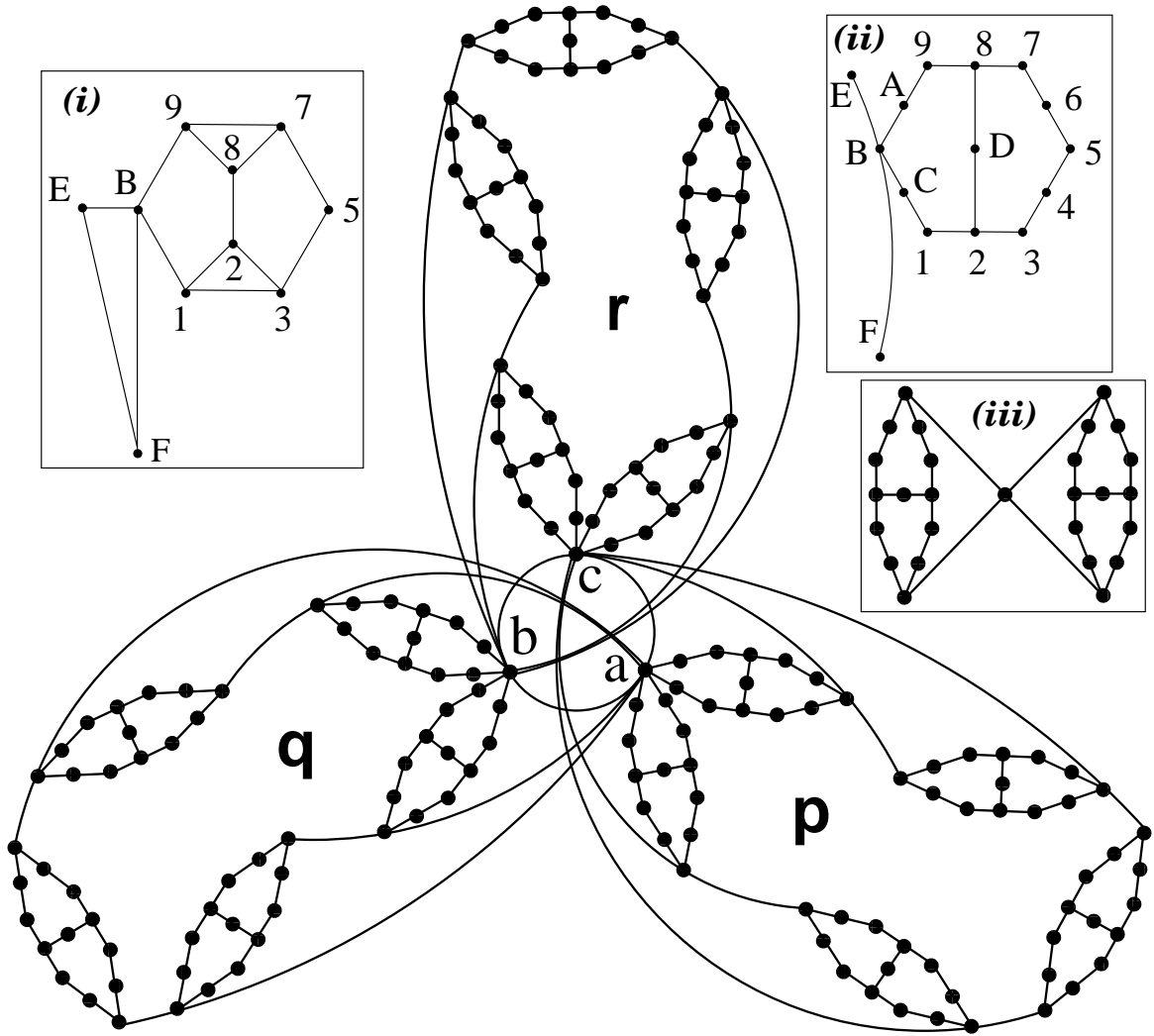}
\end{center}
\caption{Historical Kochen-Specker 192 (117) graph
\cite[Lemma 2, p.~68, Fig.~p.~69]{koch-speck} in the MMP diagram
notation. Inset {\em (i)} shows the hexagon with the adjoining
triangle from \cite[Lemma 1, p.~68]{koch-speck}. Inset {\em (ii)}
shows the same graph in our MMP diagram notation: triangles
translate as edges and everything else stays the same except that
Kochen and Specker drop the vectors that do not share edges, in
particular, vectors 4, 6, A, C and D. In
\cite[Fig.~p.~69]{koch-speck} $a=p_0$, $b=q_0$,
and $c=r_0$ hold. Here we glue these points together graphically.
The groups of hexagons $p$, $q$, and $r$ here represent the hexagons
containing $p_i$, $q_i$, and $r_i$, $i=0,\dots,4$ in
\cite[Fig.~p.~69]{koch-speck}. Inset {\em (iii)} represents
the 27(17)-point graph from \cite[Fig.~p.~70]{koch-speck} in
the MMP diagram notation.}
\label{fig:ks117}
\end{figure*}

Special attention is deserved by the first KS graph ever, given by
Kochen and Specker themselves \cite{koch-speck}. We translated it
into the MMP diagram notation in Fig.~\ref{fig:ks117},
where we also explain the correspondence between the two
notations. Vectors of the KS graph are all contained in the
three groups of five hexagons of the type shown in the inset
{\em (ii)} of the figure. Since each such hexagon contains 13
vectors and since two hexagons in each group $p$, $q$, and $r$
share a vector (vectors $a$, $b$, and $c$, respectively) this
makes 192 vectors (vertices) and 118 edges. By dropping vectors
that do not share edges Kochen and Specker obtained 117 vectors.

Let us just mention here that Kochen and Specker's 27(17)-point
graph [Fig.~\ref{fig:ks117}-{\em (iii)}] provides a partial
Boolean sub-algebra (characterising the operations of commensurable
observables in both quantum and classical mechanics) that cannot be
embedded into a Boolean algebra (i.e., not all classical
tautologies from the Boolean algebra correspond
to equalities in the partial Boolean algebra). The graph does allow
{\tt 0-1} states. It has properties similar to those of its
hexagons [cf.~Fig.~\ref{fig:drop}-(a)] since it represents a vector
system
{\tt 123,345,567,789,9AB,BC1,4DA,EFG,GHI,IJK,KLM,MNO,OPE,HQN,1RK,7RE}
with the following components:
{\tt 12\dots QR}\ =\ \{\{0,1,-2\}\{5,2,1\}\{1,-2,-1\}\{1,0,1\}\{1,1,-1\}\break
\{2,-1,1\}\hfil\{0,1,1\}\hfil\{2,1,-1\}\hfil\{1,-1,1\}\hfil\{1,0,-1\}\hfil\{1,2,1\}\hfil\{5,-2,-1\}\hfil\{0,1,0\}\hfil\{0,1,-1\}\hfil\{2,1,1\}\hfil\{1,-1,-1\}\break
\{1,1,0\}\hfil\{1,-1,2\}\hfil\{5,1,-2\}\hfil\{0,2,1\}\hfil\{5,-1,2\}\hfil\{1,1,-2\}\hfil\{1,-1,0\}\hfil\{1,1,1\}\hfil\{2,-1,-1\}\hfil\{0,0,1\}\hfil\{1,0,0\}\}\break
which does have a set of {\tt 0-1} states when
all vectors are taken into account and does not have it when the
vectors that do not share edges are dropped. However, we obviously
cannot dispense with vectors that build the system and therefore
we cannot use this system for proving the KS theorem. Of course,
Kochen and Specker were aware of this fact too and this is why
they designed the aforementioned 15-hexagon 192-vector system.

(xii) The concept of {\em KS dual diagrams}
\cite{tkadlec,tkadlec-2,svozil00} is apparently either a misnomer
or insufficiently defined.
Tkadlec claims that one arrives from a standard
3-dim KS diagram to its dual so as
to ``replace the role of points [vertices] and smooth
curves [edges]: points [vertices of the dual diagram] represent
blocks [edges of the standard diagram] and maximal smooth
blocks [maximal (?)\ edges of the dual diagram] represent
atoms [vertices of the standard diagram].''~\cite{tkadlec}
The instructions for such a construction are ambiguous, but two
figures are given in \cite{tkadlec} and \cite{tkadlec-2} and we
have tested them.
One is a dual diagram to Peres' 57 (33) diagram
\footnote[5]{\tt 123,39R,89A,47D,56E,DRE,EFG,CBD,NML,LKE,DJQ,QST,PJI,HKO,RVX,RUW,14Y,1Z5,4aA,5b8,8gB,\break
\hbox to 6.4pt{\hfill} AhF,7cH,6dI,CiO,GjP,7eM,6fS,ClN,GkT,NqX,PsV,OrU,MmU,SnV,HoX,IpW,TtW,2uB,2vF} and it reads:\break
{\tt 123,\hfil 345,\hfil 567,\hfil 869,\hfil 9AH,\hfil 8C2,\hfil 7DG,\hfil HG1,\hfil 4BA,\hfil CBD,\hfil 6gE,\hfil BhE,\hfil 3IJ,\hfil 2RO,\hfil 1VU,\hfil VPN,\hfil UML,\hfil JKN,\break
OKL,\hfil IQM,\hfil RQP,\hfil jSK,\hfil jiQ,\hfil UWX,\hfil Veb,\hfil Gfa,\hfil HCZ,\hfil ZYb,\hfil XYa,\hfil WdC,\hfil edf,\hfil TFd,\hfil TcY,\hfil 1kE,\hfil 1lj,\hfil 1mT}\break
and\hfil the\hfil other:\hfil {\tt 123,145,16C,768,7HK,4FB,GEC,89A,5IJ,HGI,EF9,KBD,JAD,CDV,KLM,\break BON,\hfil DgS,\hfil VUT,\hfil SP2,\hfil QRS,\hfil MQU,\hfil NPT,\hfil c6R,\hfil GPd,\hfil VWX,\hfil X3Y,\hfil 3Ze,\hfil EZQ,\hfil abJ,\hfil YfA,\hfil cde,\hfil ehD,\hfil acW}\break
is dual to Conway-Kochen's diagram. Of these two diagrams
only the latter does not allow {\tt 0-1} state. The former
has at least one set of {\tt 0-1} states. Then our solvers prove
that the equation system corresponding to the latter diagram does not
have a solution. Hence, neither of the two diagrams is a KS set, and
we would expect a KS dual to be a KS set.

(xiii) We obtain the class of all remaining (non-KS) vectors from ${\cal
H}^n$ by first filtering the MMP diagrams so as to keep only those that
allow {\tt 0-1} states.  Out of these, a second filter then keeps only
those diagrams whose corresponding equations have solutions.  Vectors
corresponding to these solutions are the wanted non-KS vectors.

(xiv) The presented algorithms can easily be generalised beyond the
KS theorem. One can use MMP diagrams to generate Hilbert
lattice counterexamples, partial Boolean algebras, and general
quantum algebras which could eventually serve as an algebra for
quantum computers.~\cite{mpoa99} One can also treat
any condition imposed upon inner products in ${\mathbb R}^n$ to
find solutions not by directly solving all nonlinear equations but
also by first filtering the corresponding diagrams and solving only
those equations that pass the filters.

\ack
One of us (M.~P.) would like to acknowledge the support of
the {\sl Ministry of Science, Education, and Sport} of Croatia
through the project {\it Quantum Theory of Information}, of
the Computing Centre of the Faculty of Civil Engineering
(in particular of A.~Karamati\'c who built the cluster under
OpenMosix) and of the University Computing Centre of Zagreb
through use of its clusters. He would also like to thank
Jan-\AA ke Larsson from Link\"oping University, Sweeden for
drawing his attention to Ref.~\cite{larsson}. Another
of us (B.~McK.) acknowledges support from the Australian Research
Council.

\section*{References}


\end{document}